# Real-Time Cleaning and Refinement of Facial Animation Signals

Eloïse BERSON
Dynamixyz
CentraleSupélec, CNRS, IETR,
UMR 6164, F-35000
eloise.berson@gmail.com

Catherine SOLADIE
CentraleSupélec, CNRS, IETR,
UMR 6164, F-35000
catherine.soladie@centralesupelec.fr

Nicolas STOIBER
Dynamixyz
nicolas.stoiber@dynamixyz.com

## ABSTRACT
With the increasing demand for real-time animated 3D content in the entertainment industry and beyond, performance-based animation has garnered interest among both academic and industrial communities. While recent solutions for motion-capture animation have achieved impressive results, handmade post-processing is often needed, as the generated animations often contain artifacts. Existing real-time motion capture solutions have opted for standard signal processing methods to strengthen temporal coherence of the resulting animations and remove inaccuracies. While these methods produce smooth results, they inherently filter-out part of the dynamics of facial motion, such as high frequency transient movements. In this work, we propose a real-time animation refining system that preserves -or even restores- the natural dynamics of facial motions. To do so, we leverage an off-the-shelf recurrent neural network architecture that learns proper facial dynamics patterns on clean animation data. We parametrize our system using the temporal derivatives of the signal, enabling our network to process animations at any framerate. Qualitative results show that our system is able to retrieve natural motion signals from noisy or degraded input animation.

## 1. INTRODUCTION

Motion Capture (MoCap) consists of recording the motion of a performer and transferring the animation signals to computer graphics characters. For the past two decades, motion capture has evolved into the leading technology to create realistic animation, making the process of animation content generation more reliable and accessible. It is now commonly used to produce body and facial animation in numerous applications. Despite remarkable progress made in both the motion-tracking software and hardware sensors, the output animation signals often contain artifacts due to environmental interferences, such as lighting changes, sensor noise, data occlusion, inducing reduced accuracy and jitter in the resulting animation. Post-processing is usually used to address this, typically through handmade corrections. Manual animation edition is, however, a time-consuming step, requiring highly skilled animators. For the case of facial animation, automating filtering/cleaning is a tough problem, as facial dynamics induces both low- and high-frequency of complex motions that are hard to model[1,2]. This is all the more challenging as our human eyes are experts at perceiving inconsistencies in facial motions, even the most subtle ones.

In this work, we propose a real-time facial animation cleaning system, which restores correct facial dynamics from raw, real-time motion capture input. Traditional signal processing methods such as the Kalman filter or Gaussian smoothing process often fail at preserving the subtleties of facial motions. For instance, a blink constitutes an abrupt spike in the eyelid motion signal; With aforementioned filtering frameworks, transient motion like blinks end up oversmoothed. Our approach instead learns the complex facial motion dynamics from data, and thus has the ability to preserve natural-looking motion, even transient ones. Previous works have successfully addressed the topic of learning natural motion model with neural networks, typically with CNN architectures[18, 19]. Convolutional neural network (CNN) architectures are however non-causal, as they use future time samples to process the current one, limiting their applicability to offline tasks. Causal architectures such as Recurrent Neural Networks (RNN) recently have proved successful at processing sequential data in language modeling[30], human motion prediction[12] or speech recognition[14]. In this work, with real-time applications in mind, we leverage Long Short-Term Memory (LSTM) architecture to produce natural motion models for animation filtering. Learning the dynamics of facial motion for real-time applications also differs in that it cannot rely on a known, fixed sensor frame rate. Traditional resampling algorithms are not viable, as they require to know future samples. Also, real-time source for face images, such as webcams, can have non constant framerates. We tackle this difficulty by reformulating the nature of the signal our system learns. Rather than a system predicting the next frame values given the past frame values, our recurrent network is trained to learn the values of the signal's derivatives at the current frame. Considering the temporal derivative of the motion sidesteps the problem of frame rate dependency at run-time. We use the previous estimated states and the dynamic features of the mocap-based signal as inputs of our recurrent network. In that way, our system is able to process animation with infinite length overcoming.

One tricky aspect of motion capture signal cleaning is that the 3D facial animation ground truth matching facial motion capture signals is hardly ever available; it would require a really cumbersome and expensive setup to acquire data. We overcome this difficulty by leveraging handmade animation created with a professional performance-based animation software[11], and train our network to minimize the difference between the resulting

animation and these created data. In summary, our contributions include:

- An original parametrization of the facial animation filtering problem. In particular, we define the input and the output of our recurrent network so as to free our system from frame rate dependency.
- A real-time system to clean-up a noisy animation, preserving high frequency and restoring the realistic dynamics of facial motions.

The rest of the paper is structured as follows. The following section reviews related work. In Section 3, we detail our system, the data parameterization and the training of our recurrent network. Section 4 presents results obtained with our system and comparisons with traditional filtering algorithms and non-recurrent learning-based methods. Finally, Section 5 concludes the paper with some discussion of future work.

## 2. RELATED WORK

We organize this related work in four parts: first, we review smoothing techniques used in online performance based facial animation works; Then, we detail early filtering methods derived from signal processing, followed by a review of recent learning-based methods. Finally, we examine motion editing systems.

In facial animation, real-time motion capture based animation has demonstrated high-quality results[8, 26, 36]. However, in these works the mocap-based signal is processed with smoothing techniques, losing dynamic cues of the original facial motion. Real-time performance driven facial animation systems based on depth sensors[20, 36], remove high frequency jitters using a temporal filter with exponential adaptive weights. They further improve the temporal coherence of tracking by enforcing an animation prior[9]. In the same vein, Cao et al.[8] penalize the magnitude of temporal derivatives of the output animation, resulting in smoother results. Garrido et al.[13], as well as Valgaerts et al.[33] use structure-aware regularization to improve optical flow estimation. While producing smooth results, their final animations are not free from artifacts, notably the loss of high-frequency motions in the eyes and mouth.

Properly modeling the dynamics of human motion has garnered interest among animation research for different purposes such as motion forecasting[12], motion control[17, 18], motion generation[7, 35]. Motion filtering has been a long-term research topic. Early works used standard signal processing model algorithms[6] such as Kalman Filter or exponential smoothing. Modeling the dynamics of the face is highly challenging due to the complexity and the non-linear nature of facial motions. Hence, methods using linear observation models such as the Kalman filter[5, 21, 37] appear insufficient for facial motions. Huang et al.[21] combine AAM models with Kalman filter to robustify face tracking. Existing extended nonlinear methods like particle filtering[4, 10] are often hard to tune and too memory expensive for long sequences. Many works propose to filter motion through prior-based methods, modeling motion as either dynamic, low-dimensional Gaussian Processes[32, 34], spatiotemporal bilinear model[1], Markov models[24] or as binary latent variables[31]. Although these techniques are straightforward, there would require a painstaking tuning to accurately filter every motion signal of the face. Recently, Mall et al.[27]] learn adaptive filters for each animation parameter, demonstrating successful results at cleaning any kind of actions. However, these filters are applied to the motion signals afterwards. Contrary to our method, their system is therefore non-causal. Beside, such as any temporal filter-based methods, the predicted filters depend on the framerate of the input signal.

Recently, many works on motion prediction have turned to recurrent learning-based methods. Martinez et al.[28] introduced a residual sequence-to-sequence architecture to predict short-term motion. While demonstrating state-of-the-art short term motion prediction, they fail at generating long sequences. Indeed, recurrent networks fail to cope with long sequences. In this work, we use a closed-loop architecture, feeding the output back as input to the system. In that sense, our system has knowledge of its currently generated animation output, and can continuously steer it to follow input motion capture cues. Long-term motion generation problem has been studied[12, 22, 35]. Fragkiadaki et al.[12] learn the dynamics of human motion through two architectures: a Encoder-Recurrent Decoder (ERD) and a 3 LSTM layers-based network (LSTM-3LR). Jain et al.[22] create structural RNNs to perform the same task by mixing a high-level spatio-temporal graph with the efficient sequence learning of RNNs. Both propose to handle long-term horizon forecasting by gradually adding noise to the input during the training. The noise scheduling enables their system to produce plausible motions far into the future. However, this kind of curriculum learning is hard to implement accurately. Hence, we deal with long-term motion by considering both the previous frame prediction and the actual signal. More recent works address the problem by combining recurrent networks with generative models[15, 35]. Wang et al.[35] stack a "refiner" neural network over the RNN-based generator, trained in adversarial fashion to enhance the realism of the generated motion sequence, while Habibie et al.[15] sample new motion using the variational autoencoder paradigm. These works aim at predicting future body motion rather than refining an existing animation.

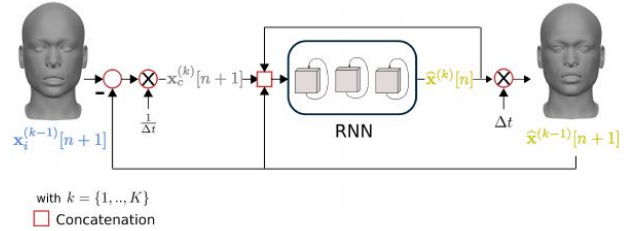

Figure 1: System overview. Our recurrent system takes as an input the first n moments of the estimate signal at time t-1 as well as a corrective moments of the inputs and regress n derivatives at time t.

This work is also related to the popular research topic of motion edition and control[15, 17, 18]. Holden et al.[17] tackle the controlled motion prediction problem using convolution neural networks. They introduce a phase functional neural network that predicts the next pose given control parameters, phase state and current pose. Recently, Berson et al.[3] propose a facial animation editing system preserving high frequency motions. While achieving impressive results, these works rely on non-causal convolutional architectures, therefore preventing usage in real-time applications. To the best of our knowledge, we are the first to provide a framerate independent cleaning system.

## 3. RNN MOTION CLEANING SYSTEM

Our goal is to enhance the accuracy and remove artifacts of a performance-based animation. To this end, we propose a learning-based solution trained to turn a noisy animation into a realistic one. In this section, we begin by detailing the particular parametrization we use to make values in our system independent of the framerate of the input data (Section 3.1). Then, we explicit the proposed architecture, as well as the training procedure of our recurrent neural network (Section 3.2). An overview of our system is shown in Figure 1.

### 3.1  Parametrization of the System

Our filtering system essentially refines motion captured facial motion to produce natural looking animation. To represent facial motion, we rely on the simple blendshapes parameterization, widely adopted throughout academia and the industry[25]. Hence, we represent a facial animation as a sequence of $N$ frames of $M = 34$ blendshapes coefficients $X = [x[0],..,x[N]]^T \in R^{NxM}$. We design our system to be framerate-independent: instead of correcting the absolute value of the current motion, we consider the normalized temporal $k$-th order derivatives of the motion signal:

$$x^{(k)}[n] = \frac{x^{(k-1)}[n+1] - x^{(k-1)}[n]}{\Delta t[n]} \quad (1)$$

where $x^{(k)}$ is the forward $k^{th}$-order derivative of the motion at the frame $n$ and $\Delta t$ is the time between two consecutive frames. With this formulation, the framerate information is factored out of the input, preventing our network to be reliant on it at both training and inference time.

At each frame $n$, our system aims at predicting the forward $k$ derivatives, $\hat{x}^{(k)}[n]$ with $k = \{1,..,K\}$ (*green* on Figure 1) given the previous estimated animation $\hat{x}[n]$, the estimated derivatives $\hat{x}^{(k)}[n$-$1]$, and the current *corrective* forward derivatives $x_c^{(k)}[n]$ (*grey* on Figure 1, see below for details). Finally, from $\hat{x}^{(k)}[n]$, we recover the estimated $K$-$1$ derivatives using the equation 1 (i=0 corresponds to the absolute blendshape values).

In this study, we have observed that feeding back to the system a measure of how much the currently produced state deviates from the real input signal improves the performance of our system. Hence, we give as input to our network the $k$ corrective temporal derivative, $x_c^{(k)}$, of the input signal, $x_i$ formulated as:

$$x_c^{(k)}[n] = \frac{x_i^{(k-1)}[n+1] - \hat{x}^{(k-1)}[n]}{\Delta t[n]} \quad (2)$$

where $X_i$ and $\hat{X}$ are respectively the input and the generated animation. Besides, we add prior model knowledge about the dynamics of the sequence by adding a residual connection between the input and the output of each RNN cell of our network as Martinez et al.[28].

### 3.2  Architecture and Training Details of Our Recurrent Network

Our approach is based on a recurrent network[12, 28]. The goal of this work is to learn to generate proper facial dynamics from data. We benefit from well-established LSTM[16] capacities to model and forget temporal dependencies to carry out this task. Our network is depicted in Figure 1 and mainly consists of a sequence of LSTM layers with a stacked final dense output layer to get dimensions matching the output features. As our network is thought for real-time animation, it is inputted with past time samples. Its objective is to predict a plausible estimation of facial motion given previously estimated states and the corrective derivative of the input signal (Equation 2). Therefore, at training time, we formulate the cost function as the mean square error (MSE) between the animation made by an artist, $X_{gt}$ and the system's estimate output state $\hat{X}$ : $L_{MSE} = \| X_{gt} - \hat{X} \|^2$. We also encourage our network to focus on the higher-order dynamics of facial motion with an MSE between the derivatives of the estimate motion and the ground truth one: $L_{der} = \sum_{k=1}^{K} \|x_{gt}^{(k)} - \hat{x}^{(k)}\|^2$.

As Berson et al.[3], we add a loss $L_{dis}$, to focus preservation of some key inter-vertices distances between the estimate and the ground truth animations: $L_{dis} = |D_{gt} - \hat{D}| + \alpha_{dis}|D_i - \hat{D}|$, where $D \in R^{nx6}$ is a sequence of intervertices distances derived on the same mesh animated with the ground truth and the estimated blendshapes coefficients. In this loss, we include six distances: the first three one measures the extent between the upper and the lower lips (at the middle and at one and two third of the mouth), the fourth is between the mouth corners and the last ones between the right and left eyelids. This loss emphasizes the salient role of the lips and eyes to convey expressivity and communicational cues in facial animation. Finally, we optimize the following cost function: $L = L_{MSE} + w_{der}L_{der} + w_{dis}L_{dis}$.

For all our experiments, we set $w_{der}$ and $w_{dis}$ at 0.01 and 0.1, $\alpha_{dis}$ at 0.8 and use $K$ =1. The dimension of the hidden states is set to 128 for every LSTM layer. Our network is optimized using the ADAM algorithm[23]. During the training, we add a dropout[29] of 0.3 to avoid overfitting. We set the initial learning rate at 0.001.

## 4. RESULTS

In this section, after introducing the dataset, we demonstrate the capacity of our model to clean-up a noisy performance-based animation while preserving a plausible facial motion dynamic. We also compare our system to standard signal processing methods to highlight the difficult task of hyperparameters tuning in the case of motion signals filtering. Finally, we demonstrate the relevance of our recurrent structure by comparing our system with non-recurrent learning methods.

### 4.1  Data Gathering

To train our network, we use as an input, sequences of animation generated by an automatic face tracking solution, without any post processing. We use the corresponding handmade animations as the groundtruth. Our training set contains 52 sequences recorded at different framerates between 30 and 120 frames-per-second (fps). Our test set is made of 4 sequences recorded at 60 fps. In total, we gather around 285 000 frames for the training set and 5000 frames for the test set, representing around 49 minutes of animation. We divide every sequence in chunks of 200 frames with an overlap of 150 frames. As the network can learn both short-term and long-terms temporal patterns, larger chunks improve long-term processing.

### 4.2  Motion Refinement

Facial performance-base systems often rely on video sources to capture motion and solve for animation. Most of the time, either due to sensor quality or environmental factors (lighting changes, occlusions), the delivered animation contains noise and inaccuracies. Some crucial properties of facial animation, such as the amplitude of the movements, are often lost resulting in a less expressive animation. As shown in Figure 2, our system recovers the natural dynamic of the facial motion. For instance, providing the previous estimated velocities, our learning-based method detects an upcoming blink and corrects the animation to get a natural full closure of the eyes. Our system has to predict the expression parameters of the whole face at each time step. Hence, it learns the correlation between animation parameters, as we can see in Figure 3 (top). Our system can "magnify" the motion by augmenting the protrusion movements in the animation, which were absent from the noisy input. Conversely, when unrealistic blendshapes activation patterns appear, our system efficiently smooths the signal (see Figure 3 (middle, bottom)).

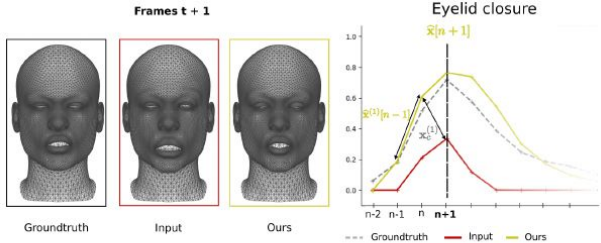

Figure 2: Our system detects a blink pattern and correct the motion to retrieve a realistic full closure of the eyelids.

The time to infer one frame is less than 0.5 ms on GPU (GeForce GTX 1060). Hence our system could be integrated in any real-time facial animation software. More results on full animations are provided at https://youtu.be/yAOGq18IU4k.

## 4.3 Comparison with Temporal Filters

One shortcoming of standard filtering methods is hyperparameters tuning. One needs to find a trade-off between preserving high-frequency patterns such as a blink or noise or getting a smooth animation and losing the natural dynamics of some part of the face. For instance, one popular filtering algorithm for real-time processing is exponential smoothing: $\hat{x}[n] = \gamma x[n] + (1 - \gamma)\hat{x}[n-1]$. Setting a high $\gamma$ results in an estimate signal, which is more faithful to the input. In this case, more subtle motion patterns of the input signal are kept. Conversely, setting a low $\sigma$ prevents from high variations in the estimated motion signal leading to a smoother animation. Another popular smoothing scheme is Gaussian-based filtering. It consists of convolving the input signal with a Gaussian window, and thus requires having the whole signal. The smoothness of the output signal depends on the resolution of the window fixed by standard deviation $\sigma$ of the Gaussian. The lower the standard deviation, the higher the temporal resolution of the window. A narrow window better preserves the fine temporal details. In both cases, the techniques only filter the animation and cannot refine it. The dynamics of the different parts of the face are very different and complex to model. While the eyelids motion is composed mainly of flat portions and quick spikes corresponding to blinks, frowning movements consist of more subtle variations with variable lengths. Handmade tuning of $\gamma$ or $\sigma$ parameters is thus a cumbersome task. By learning the inherent dynamic faces, our method is free from such frequency parameters tuning.

We compare our system with the temporal smoothing algorithm parametrized with two different values of $\gamma$ 0.3 and 0.5 (see Figure 4a) and with the Gaussian smoothing using a window with $\sigma$ of 1.0 and 5.0 (see Figure 4b). Our system is able to enhance eyelids signal producing accurate closures of the eyes. Indeed, our system detects the inaccurate spikes observed on the "*Top right lip sneer*" blendshapes and corrects it to produce a smoother and a more natural motion signal. Conversely, both methods process the spikes observed on the "*Top right lip sneer*" blendshapes and on the "*Eyelid*" motion similarly, by either preserving it or smoothing it. At each time step, our system is fed with the motion of the whole face. As shown in Figure 4, our system is able to learn natural correlations in facial motion and use this knowledge to correct and generate more accurate motion sequences, even if the input tracking is inexact. We also numerically compare the MSE of those algorithms on the test set in Table 1. Our system gets the lowest MSE, while smoothing methods get a MSE similar to the MSE between the input and the ground truth.

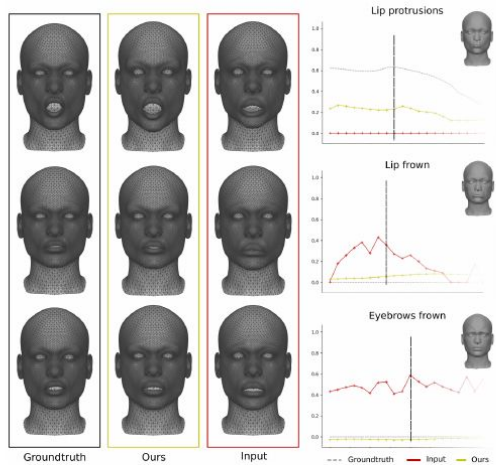

Figure 3: Our system corrects the motion of every part of face: either by increasing the motion such as the "lip pro-trusion" motion or by smoothing the lips or the eyes frown movement.

Table 1. Quantitative comparison with non-recurrent methods.

|  | MSE |
| --- | --- |
| Exponential ($\gamma$ :0.3) | 0.0170 |
| Exponential ($\gamma$ :0.8) | 0.0461 |
| Gaussian ($\sigma$ :1) | 0.0170 |
| Gaussian ($\sigma$ :5) | 0.0167 |
| Raw input | 0.0173 |
| Ours | **0.0140** |

## 4.4 Comparison with Non Recurrent Learning Methods

We compare our system with non-recurrent learning methods: a Fully Connected neural network (FC) and a non-neural machine-learning algorithm using Gradient-Boosted Trees (GBT). In these algorithms, the estimate of previous derivatives is not fed back at train time. Hence, we adapt the input parameterization by replacing the estimate of previous outputs with parameters of the input signal. At each frame $n$, we estimate the $K$ derivatives of the corrected signal at time $n + 1$, given the $K$ derivatives of the input animation at $n$ and $n - 1$, and the current state of the input.

As these methods are not recurrent and to avoid accumulating errors through time, the estimate animation is derived as:
$X_r^{(k)} = \alpha_r(\Delta t \hat{X}^{(k+1)} + \hat{X}_r^{(k)}) + (1 - \alpha_r)(\hat{X}^{(k)})$,

with $\hat{X}^{(0)} = X_i$ and $X_r^{(K)} = \hat{X}_r^{(k)}$. $X_r$ is the resulting animation, while $\hat{X}$ is the estimated animation by the FC/GBT algorithms. We augment the training set by upsampling and downsampling each sequence with a factor 2 to avoid overfitting. We train both the GBT and FC using $K=4$ and produce the final animation using empirically chosen values of $\alpha_r=0.97$ (GBT) and $\alpha_r=0.9$ (FC). During the training, we optimize the loss $L_{der}$. We also test these algorithms on animations with different framerates. Figure 5 depicts two frames extracted from the resulting animation of GBT or FC and the animation estimated by our system when fed with a performance-based animation recorded at 30 fps. Compared to these algorithms, our system not only rectifies the motion signals but also enhances the expressiveness of the animation. As we can see, non-recurrent methods tend to flatten the motion signals, whereas our system produces natural motion patterns such as the eyebrows frowning or the protrusion of the lips. We also numerically compare the MSE error obtained on the test set, and observe that our recurrent method gets a lower MSE than GBT or FC architectures (Table 2). Dynamic animation results are shown in the video available at https://youtu.be/yAOGq18IU4k .

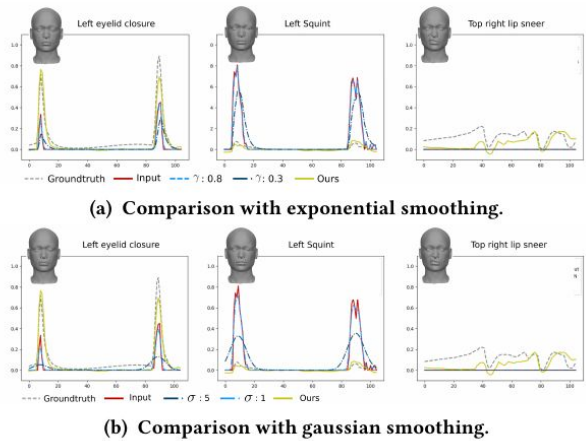

(a) Comparison with exponential smoothing.

(b) Comparison with gaussian smoothing.

Figure 4: Common filtering methods require hyperparameter tuning to balance between oversmoothing and details preservation. Our system learns the dynamic of facial motions so as to tailor the filtering process for each motion. Hence, it can enhance a blink motion while removing unrealistic spikes in "left squint" motion. It is also able to retrieve natural patterns in the "right lip sneer" motion for instance.

**Table 2. Quantitative comparison with non-recurrent methods.**

|  | MSE |
|---|---|
| GBT | 0.0451 |
| FC | 0.0461 |
| Ours | **0.0140** |

## 5. CONCLUSION
In this work, we develop a facial animation cleaning and refinement system. Taking blendshape animation as input, such as raw motion-capture animation, our system successfully filters and enhances the animation, in real-time, regardless of the input framerate. Contrary to traditional signal processing methods, our system learns the dynamics of facial motion on realistic data, in order to be capable of removing noise at all frequencies and yet preserves high-frequency transient motions. We also compare our system with non-recurrent learning-based methods to highlight the relevance of our recurrent neural network architecture. Besides, as demonstrated on mocap-based animation, our system refines the animation by delivering natural motion patterns and realistic correlations between different parts of the face. By learning the derivatives of the motion rather than the motion's absolute values, we free our system from framerate dependency, enabling it to process any input animation in real-time. As any learning-based system, our system strongly depends on the data used during the training. Thus, our system tends to produce only motion patterns that it has seen. Thus, more data would be strongly valuable to improve our results. Also, our system refines animation in the style of the animations it has learned on. It is very likely that the amount and the type of data (inevitably bearing the style of the animators who made it), strongly affects the delivered animation. A more in-depth analysis on database dependency and style learning would be required to get full control of the animation quality that our system outputs.

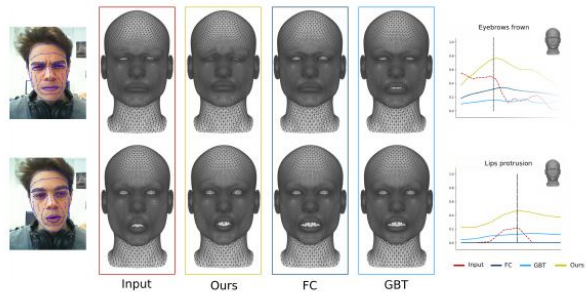

Figure 5: Comparison with non-recurrent machine-learning algorithms: a Gradient Boosting Trees (GBT) and a Fully Connected neural network (FC). Our system not only rectifies the motion signals but also enhances the expressiveness of the animation, such as enforcing the eyebrows frowning movements or the lip protrusion.

## 6. REFERENCES
[1] Akhter, I., Simon, T., Khan, S., Matthews, I. and Sheikh, Y. 2012. Bilinear spatiotemporal basis models. *ACM Transactions on Graphics*. 31, 2 (Apr. 2012), 1–12.

[2] Bermano, A.H., Bickel, B., Gross, M., Bradley, D., Beeler, T., Zund, F., Nowrouzezahrai, D., Baran, I., Sorkine-Hornung, O., Pfister, H. and Sumner, R.W. 2014. Facial performance enhancement using dynamic shape space analysis. *ACM Transactions on Graphics*. 33, 2 (Apr. 2014), 1–12.

[3] Berson, E., Soladié, C., Barrielle, V. and Stoiber, N. 2019. A Robust Interactive Facial Animation Editing System. *Proceedings of the 12th Annual International Conference on Motion, Interaction, and Games* (New York, NY, USA, 2019), 26:1–26:10.

[4] Bray, M., Koller-Meier, E. and Van Gool, L. 2007. Smart particle filtering for high-dimensional tracking. *Computer Vision and Image Understanding*. 106, 1 (2007), 116–129.

[5] Bregler, C. 1997. Learning and recognizing human dynamics in video sequences. *Proceedings of IEEE Computer Society Conference on Computer Vision and Pattern Recognition* (San Juan, Puerto Rico, 1997), 568–574.

[6] Bruderlin, A. and Williams, L. 1995. Motion signal processing. *Proceedings of the 22nd annual conference on Computer graphics and interactive techniques - SIGGRAPH '95* (Not Known, 1995), 97–104.

[7] Bütepage, J., Black, M., Kragic, D. and Kjellström, H. 2017. Deep representation learning for human motion prediction and classification. *arXiv:1702.07486 [cs]*. (Apr. 2017).

[8] Cao, C., Hou, Q. and Zhou, K. 2014. Displaced dynamic expression regression for real-time facial tracking and animation. *ACM Transactions on Graphics*. 33, 4 (Jul. 2014), 1–10.

[9] Cao, C., Weng, Y., Lin, S. and Zhou, K. 2013. 3D shape regression for real-time facial animation. *ACM Transactions on Graphics (TOG)*. 32, 4 (2013), 41.

[10] Doucet, A., De Freitas, N. and Gordon, N. 2001. An introduction to sequential Monte Carlo methods. *Sequential Monte Carlo methods in practice*. Springer. 3–14.

[11] Dynamixyz 2019. *Performer*.

[12] Fragkiadaki, K., Levine, S., Felsen, P. and Malik, J. 2015. Recurrent network models for human dynamics. *Proceedings of the IEEE International Conference on Computer Vision* (2015), 4346–4354.

[13] Garrido, P., Valgaert, L., Wu, C. and Theobalt, C. 2013. Reconstructing detailed dynamic face geometry from monocular video. *ACM Transactions on Graphics*. 32, 6 (Nov. 2013), 1–10.

[14] Graves, A. and Jaitly, N. 2014. Towards end-to-end speech recognition with recurrent neural networks. *International conference on machine learning* (2014), 1764–1772.

[15] Habibie, I., Holden, D., Schwarz, J., Yearsley, J., Komura, T., Saito, J., Kusajima, I., Zhao, X., Choi, M.-G. and Hu, R. 2017. A Recurrent Variational Autoencoder for Human Motion Synthesis. *IEEE Computer Graphics and Applications*. 37, (2017), 4.

[16] Hochreiter, S. and Schmidhuber, J. 1997. Long short-term memory. *Neural computation*. 9, 8 (1997), 1735–1780.

[17] Holden, D., Komura, T. and Saito, J. 2017. Phase-functioned neural networks for character control. *ACM Transactions on Graphics*. 36, 4 (Jul. 2017), 1–13.

[18] Holden, D., Saito, J. and Komura, T. 2016. A deep learning framework for character motion synthesis and editing. *ACM Transactions on Graphics*. 35, 4 (Jul. 2016), 1–11.

[19] Holden, D., Saito, J., Komura, T. and Joyce, T. 2015. Learning motion manifolds with convolutional autoencoders. (2015), 1–4.

[20] Hsieh, P.-L., Ma, C., Yu, J. and Li, H. 2015. Unconstrained realtime facial performance capture. *Proceedings of the IEEE Conference on Computer Vision and Pattern Recognition* (2015), 1675–1683.

[21] Huang, C., Ding, X. and Fang, C. 2012. Pose robust face tracking by combining view-based AAMs and temporal filters. *Computer Vision and Image Understanding*. 116, 7 (2012), 777–792.

[22] Jain, A., Zamir, A.R., Savarese, S. and Saxena, A. 2016. Structural-RNN: Deep learning on spatio-temporal graphs. *Proceedings of the IEEE Conference on Computer Vision and Pattern Recognition* (2016), 5308–5317.

[23] Kingma, D.P. and Ba, J. 2014. Adam: A Method for Stochastic Optimization. *arXiv:1412.6980 [cs]*. (Dec. 2014).

[24] Lehrmann, A.M., Gehler, P.V. and Nowozin, S. 2014. Efficient Nonlinear Markov Models for Human Motion. *2014 IEEE Conference on Computer Vision and Pattern Recognition* (Columbus, OH, USA, Jun. 2014), 1314–1321.

[25] Lewis, J.P., Anjyo, K., Rhee, T., Zhang, M., Pighin, F. and Deng, Z. 2014. Practice and Theory of Blendshape Facial Models. *Eurographics (State of the Art Reports)*. 1, 8 (2014), 2.

[26] Li, H., Yu, J., Ye, Y. and Bregler, C. 2013. Realtime Facial Animation with On-the-fly Correctives. *ACM Trans. Graph.* 32, 4 (2013), 42:1–42:10.

[27] Mall, U., Lal, G.R., Chaudhuri, S. and Chaudhuri, P. 2017. A Deep Recurrent Framework for Cleaning Motion Capture Data. *arXiv:1712.03380 [cs]*. (Dec. 2017).

[28] Martinez, J., Black, M.J. and Romero, J. 2017. On human motion prediction using recurrent neural networks. *arXiv:1705.02445 [cs]*. (May 2017).

[29] Srivastava, N., Hinton, G., Krizhevsky, A., Sutskever, I. and Salakhutdinov, R. 2014. Dropout: a simple way to prevent neural networks from overfitting. *The journal of machine learning research*. 15, 1 (2014), 1929–1958.

[30] Sutskever, I., Martens, J. and Hinton, G.E. 2011. Generating text with recurrent neural networks. *Proceedings of the 28th international conference on machine learning (ICML-11)* (2011), 1017–1024.

[31] Taylor, G.W., Hinton, G.E. and Roweis, S.T. 2007. Modeling human motion using binary latent variables. *Advances in neural information processing systems* (2007), 1345–1352.

[32] Urtasun, R., Fleet, D.J., Geiger, A., Popović, J., Darrell, T.J. and Lawrence, N.D. 2008. Topologically-constrained latent variable models. *Proceedings of the 25th international conference on Machine learning - ICML '08* (Helsinki, Finland, 2008), 1080–1087.

[33] Valgaerts, L., Wu, C., Bruhn, A., Seidel, H.-P. and Theobalt, C. 2012. Lightweight binocular facial performance capture under uncontrolled lighting. *ACM Transactions on Graphics*. 31, 6 (Nov. 2012), 1.

[34] Wang, J.M., Fleet, D.J. and Hertzmann, A. 2008. Gaussian Process Dynamical Models for Human Motion. *IEEE Transactions on Pattern Analysis and Machine Intelligence*. 30, 2 (Feb. 2008), 283–298.

[35] Wang, Z., Chai, J. and Xia, S. 2019. Combining recurrent neural networks and adversarial training for human motion synthesis and control. *IEEE transactions on visualization and computer graphics*. (2019).

[36] Weise, T., Bouaziz, S., Li, H. and Pauly, M. 2011. Realtime performance-based facial animation. *ACM transactions on graphics (TOG)* (2011), 77.

[37] Weng, S.-K., Kuo, C.-M. and Tu, S.-K. 2006. Video object tracking using adaptive Kalman filter. *Journal of Visual Communication and Image Representation*. 17, 6 (2006), 1190–12.